%% file: neurips_mlsb_2020.tex
\documentclass{article}

\usepackage[final, nonatbib]{neurips_mlsb_2020}

\usepackage[utf8]{inputenc} 
\usepackage[T1]{fontenc}    
\usepackage{hyperref}       
\usepackage{url}            
\usepackage{booktabs}       
\usepackage{amsfonts}       
\usepackage{nicefrac}       
\usepackage{microtype}      
\usepackage{graphicx}       

\title{Pre-training Protein Language Models with Label-Agnostic Binding Pairs Enhances Performance in Downstream Tasks}

\author{%
  Modestas Filipavicius \\
  ETH Zürich \\
  Zürich, Switzerland \\
  \texttt{mfilipav@ethz.ch} \\

  \AND
    Matteo Manica \\
    IBM Research Zürich \\
    Rüschlikon, Switzerland \\
    \texttt{tte@zurich.ibm.com} \\
    \And
    Joris Cadow \\
    IBM Research Zürich \\
    Rüschlikon, Switzerland \\
    \texttt{dow@zurich.ibm.com} \\
    \And
    Maria Rodriguez Martinez \\
    IBM Research Zürich \\
    Rüschlikon, Switzerland \\
    \texttt{mrm@zurich.ibm.com} \\
}

\begin{document}

\maketitle

\begin{abstract}
Less than 1\% of protein sequences are structurally and functionally annotated. Natural Language Processing (NLP) community has recently embraced self-supervised learning as a powerful approach to learn representations from unlabeled text, in large part due to the attention-based context-aware Transformer models.
In this work we present a modification to the RoBERTa model by inputting during pre-training a mixture of binding and non-binding protein sequences (from STRING database). However, the sequence pairs have no label to indicate their binding status, as the model relies solely on Masked Language Modeling (MLM) objective during pre-training.
After fine-tuning, such approach surpasses models trained on single protein sequences for protein-protein binding prediction, TCR-epitope binding prediction, cellular-localization and remote homology classification tasks. We suggest that the Transformer's attention mechanism contributes to protein binding site discovery.
Furthermore, we compress protein sequences by 64\% with the Byte Pair Encoding (BPE) vocabulary consisting of 10K subwords, each around 3-4 amino acids long.
Finally, to expand the model input space to even larger proteins and multi-protein assemblies, we pre-train Longformer models that support 2,048 tokens.
Further work in token-level classification for secondary structure prediction is needed. Code available at: \url{https://github.com/PaccMann/paccmann_proteomics}
\end{abstract}

\section{Introduction}
Inferring protein properties from the primary amino acid sequence is of paramount importance in the light of sequencing technology advances resulting in a vast number of proteins with unknown characteristics. UniProt database, as of \texttt{2020\_011} release, contains 180M protein sequences, and only 560K of them (or 0.31\%) are labeled \cite{uniprot2019uniprot}.
The few proteins that are functionally and structurally characterized belong to a highly exclusive group of families with immediate industrial and therapeutic potential, such as GFPs, viral Env proteins or more recently CAS nucleases. Therefore, most of proteins stay in the dark matter of the proteome -- especially those from Prokarya and viruses, or intrinsically disordered or post-translationally modified proteins \cite{ross2016dark}.

As early as 1970s Afinsen proposed that the protein structure and function is encoded in its primary sequence \cite{anfinsen1973principles}. Today, the prevailing hypothesis claims that under evolutionary forces protein sequence space was sampled to form a closed set of structural and functional motifs, since there was a strong evolutionary pressure to reuse and recycle these components \cite{alquraishi2019end}. Thus, even the unlabeled raw protein sequences should be enough to implicitly understand the language of proteins.

Recently, self-supervised learning established itself as a powerful method for learning useful information from unlabeled sequences, especially attention-based Transformer models \cite{vaswani2017attention}. They work by first pre-training a large context-aware attention model on millions of unlabeled text lines with a proxy task such as predicting the next word in a sentence given all previous words \cite{peters2018deep, radford2018improving, radford2019language} or predicting masked-out words from their context \cite{devlin2018bert, liu2019roberta}. In the second step, the model is fine-tuned specifically for each downstream task in a supervised manner with labeled datasets. 

What if we treated protein sequences as sentences and individual (or several continuous) amino acids as words? Throughout 2019 and 2020 we saw a surge of such NLP-based techniques applied to protein sequences, showing preliminary potential to extract useful biological information from massive unlabeled datasets \cite{rives2019biological, rao2019evaluating, min2019pre, strodthoff2019udsmprot, alley2019unified, bepler2019learning, heinzinger2019modeling, elnaggar2020prottrans, nambiar2020transforming}.

\paragraph{Our contributions to transformer-based protein language modeling:}
\begin{itemize}
    \item Demonstrate that pre-training a RoBERTa language model, solely with the masked language modeling (MLM) objective, on a mixture of binding and random protein pairs, as opposed to single sequences, results in superior downstream protein classification performance.
    
    \item We compress the protein sequence space by 64\% by expanding the 20 amino acid character vocabulary to 10K subword tokens by Byte-Pair Encoding (BPE) algorithm \cite{sennrich2015neural}.

    \item Prepare new pre-training and protein-protein binding prediction datasets from STRING database \cite{szklarczyk2019string}.
    
    \item Pre-train a Longformer LM \cite{beltagy2020longformer} with a maximum sequence length of up to 2,048 tokens, accommodating long protein sequences or multi-protein assemblies.
\end{itemize}

\section{Related work}
UniRep was the first study to apply deep learning to obtain protein representations from unlabeled sequences \cite{alley2019unified}. In autoregressive manner a multiplicative LSTM was pre-trained on 22M Pfam \cite{el2019pfam} sequences to learn a 1,900-dim representation for each amino acid. Next, Bepler \& Berger (2019) successfully predicted the secondary structure by jointly pre-training bidirectional LSTM on protein sequence pairs, and simultaneously supervising the model by predicting the structure similarity between those sequences and amino acid contacts within a sequence. Inspired by ULMFit's \cite{howard2018universal} "pre-train first then fine-tune" procedure, UDSMProt \cite{strodthoff2019udsmprot} pre-trained a first protein transfer learning model with a multi-layer LSTM.

Recent protein language modelling efforts used the context-dependant LM architectures: ELMo and BERT. ELMo \cite{peters2018deep} has been applied in P-ELMO \cite{bepler2019learning}, SeqVec \cite{heinzinger2019modeling}, PLUS-RNN \cite{min2019pre}. ELMo learns context-dependent representations by predicting the next token separately in forward and reverse directions with LSTM \cite{hochreiter1997long}. Later these two representations are combined, however they struggle when bidirectional context is needed. BERT \cite{devlin2018bert} solved this problem by pre-training Transformer encoder units with masked language modeling (MLM) objective. This architecture was used for proteins by ESM \cite{rives2019biological}, TAPE \cite{rao2019evaluating} and ProtTrans \cite{elnaggar2020prottrans} studies.

In parallel to our work, ProBERTa \cite{nambiar2020transforming} was also trained with a BPE subword representation. We coincidentally chose the same size for BPE vocabulary (10K) and trained with RoBERTa. However, our model is different is two key aspects. First, we pre-trained on much bigger datasets with at least 10M sequences, as opposed to 0.5M sequences from SwissProt. Second, our model was bigger with 12 layers and 12 heads. Together this implies that a bigger model trained on larger datasets albeit for a few epochs achieves better performance than a shallow and narrow model trained on small datasets for many steps \cite{li2020train, gururangan2020don}.

In terms of accommodating long protein sequences, ProtBert \cite{elnaggar2020prottrans} trained Transformer-XL and XLNet atoregressive models which support long sequences by re-using hidden states of the previous sub-sequences. Conversely, our Longformer \cite{beltagy2020longformer} is an auto-encoding LM which learns bidirectional contexts which is better suited for sequence pair tasks like binding prediction.

\section{Methods}
\subsection{Pre-training procedure}
We tested two strategies of providing protein sequences to the Transformer architecture: \textit{single protein sequences} vs \textit{paired sequences}. The two sequences in the pair were either strong binders (as reported in the STRING DB \cite{szklarczyk2019string}) or randomly matched. The single sequence models are referred to as  \textit{Pfam, String, StringLF} and \textit{SwissProt}, while paired sequence models are called \textit{String2Seq} and \textit{StringLF2Seq} (we did not prepare two-sequence models for Pfam and SwissProt).\footnote{Pretrained models and data available at: \url{https://ibm.ent.box.com/v/paccmann-proteomics-data}}

\begin{figure}
\centering
\includegraphics[width=1.1\linewidth]{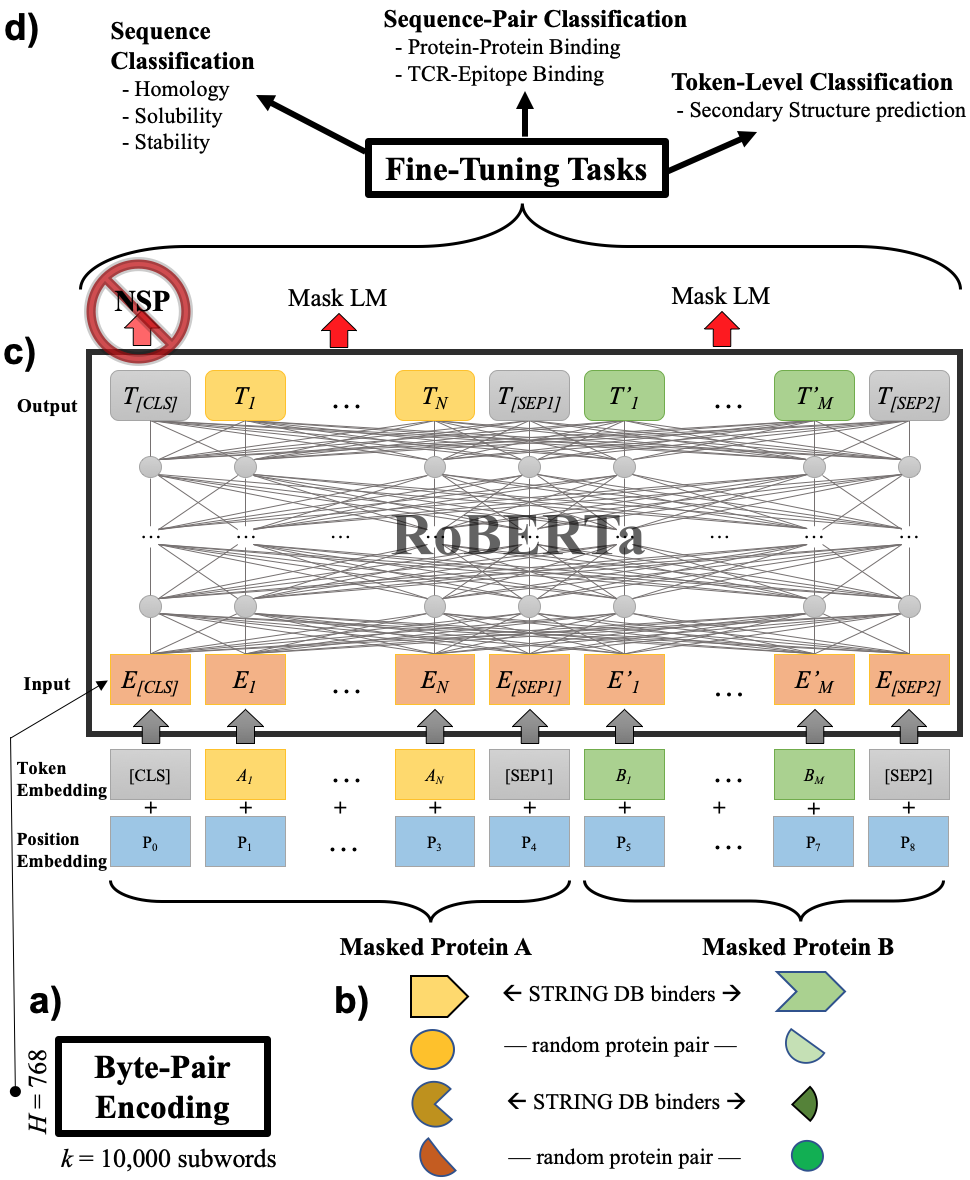}
\caption{Model is pre-trained by a mixture of binding and non-binding protein sequences from STRING DB (b), using only the MLM objective. Byte-pair encoding with a 10k subword vocabulary (a) enables inputting 64\% longer protein sequences compared to character level embedding. After pre-training, the model can be fine-tuned for 3 types of protein prediction tasks (d). $E_i$ and $T_i$ represent input and contextual embeddings for token $i$. \texttt{CLS} is a special token for classification-task output, while \texttt{SEP} separates two non-consecutive sequences.
}
\label{fig:architecture}
\end{figure}

Our pre-training architecture is a derivative of RoBERTa \cite{liu2019roberta}, and is depicted in \textbf{Figure \ref{fig:architecture}} with a paired sequences input. For more information about the architecture, consult \cite{vaswani2017attention, devlin2018bert, liu2019roberta}. We pre-train only with the masked language modeling (\textbf{MLM}) objective, by masking out 15\% of input tokens at random and reporting cross-entropy (perplexity) values for their binary prediction. 

Models are tested on the Pfam holdout dataset, which was used in the TAPE study \cite{rao2019evaluating} and consists of 44K protein sequences from 6 Pfam families - PF01112, PF03417, PF03577, PF14604, PF18346, PF18697. As a reference, we use TAPE study, where their BERT model (with no NSP objective) achieved CE 10.0. However, their model was trained with 25 tokens, thus the maximum CE value could have been 25. In our case, it is 10K.
All models had a deep ($L=12$ layers) and wide ($H=12$ attention heads) architecture, except for SwissProt ($L=6$) 
The remaining pre-training parameters are displayed in \textbf{Table \ref{tab:pretraining_hyperparams}}.
We used four Tesla P100 SXM2 GPUs with 16Gb memory, ensuring to put $2^{16}$ tokens per effective batch, while the mixed precision training (fp16) \cite{nvidia_apex} reduced memory consumption and sped-up training.
All of our protein language models are pre-trained and fine-tuned using RoBERTa architecture implemented in \texttt{Transformers} library \cite{wolf2019transformers} and \texttt{PyTorch 1.3.1} \cite{paszke2019pytorch}.

To accommodate protein sequences of 2,048 tokens long, we used the \textbf{Longformer} architecture \cite{beltagy2020longformer}, a derivative RoBERTa which utilizes a predefined length attention window $w$ within which the expensive quadratic complexity self-attention is computed. We follow implementation from \cite{beltagy2020longformer}, with an attention window of 512 tokens. Depending on the task, some tokens are assigned global attention. For instance, in sequence classification tasks the \texttt{CLS} token pools attention from all sequence tokens.
Since pre-training from scratch is computationally expensive, we start with RoBERTa checkpoints from earlier experiments trained on up 512 token-long sequences. Initially training with the shorter sequences, and towards the end with expensive longer ones is a common practice \cite{devlin2018bert, beltagy2020longformer}.

\subsection{Input sequence representation with Byte-Pair Encoding (BPE)}
BPE \cite{sennrich2015neural} is a sub-word segmentation algorithm commonly used in Neural Machine Translation, and more recently in GPT2 \cite{radford2019language} and RoBERTa \cite{liu2019roberta} LMs. BPE, in a deterministic manner, creates a vocabulary from initial subwords by merging the most frequently occurring subword pairs, until the pre-defined vocabulary size ($k$) is reached. In our case, $k=10,000$ subwords.

\subsection{Pre-training data}
\input{tables/pretraining_results}
We pre-trained on protein sequences from SwissProt \cite{uniprot2019uniprot}, Pfam \cite{el2019pfam} and STRING \cite{szklarczyk2019string} databases, for more information see \textbf{Table \ref{tab:pretraining_results}}. Most notably, we prepared two pre-training datasets which contain protein sequence pairs from STRING, half of them are strong binders (binding interaction score is at least 700, in 0-1,000 confidence scale), while the other half are random pairs. \textit{String2Seq} model has two binding pairs that together are shorter than 512 BPE token limit, while \textit{StringLF2Seq} model accommodates two sequences which together are up to 2,048 tokens-long.

\subsection{Fine-tuning tasks}
\input{tables/data_all_finetune_tasks}

Pre-training can be assessed with cross-entropy loss on MLM objective, but the ultimate assessment comes from the downstream performance on protein prediction tasks. 
As shown in \textbf{Figure \ref{fig:architecture}D}, we have attempted three types of tasks . First, for single protein sequence classification, we plug in task-specific inputs into the pre-trained model, with a single modification of a supplementary output layer which takes in the representation from \texttt{CLS} to give class SoftMax probabilities. Similarly, for protein sequence pair classification, the relationship between two sequences (separated by a \texttt{SEP} token) is learned by the \texttt{CLS} token. Finally, for token level classification (sequence annotation), all tokens except \texttt{CLS} pass through an extra SoftMax layer to give a class probability for a token.
See \textbf{Table \ref{tab:data_all_finetune_tasks}} and \textbf{Appendix \ref{seq:appendix_finetune_datasets}} for fine-tuning data description and hyperparameters. Parameter search was performed for learning rate $\{1e-5, 3e-5, 5e-5\}$, batch size $\{8, 16, 32\}$ and learning rate weight decay $\{0.05, 0.1\}$.

\section{Results}
\subsection{BPE allows efficient sequence compression for sequence classification tasks}
\begin{figure}[t]
\centering
\includegraphics[width=0.85\linewidth]{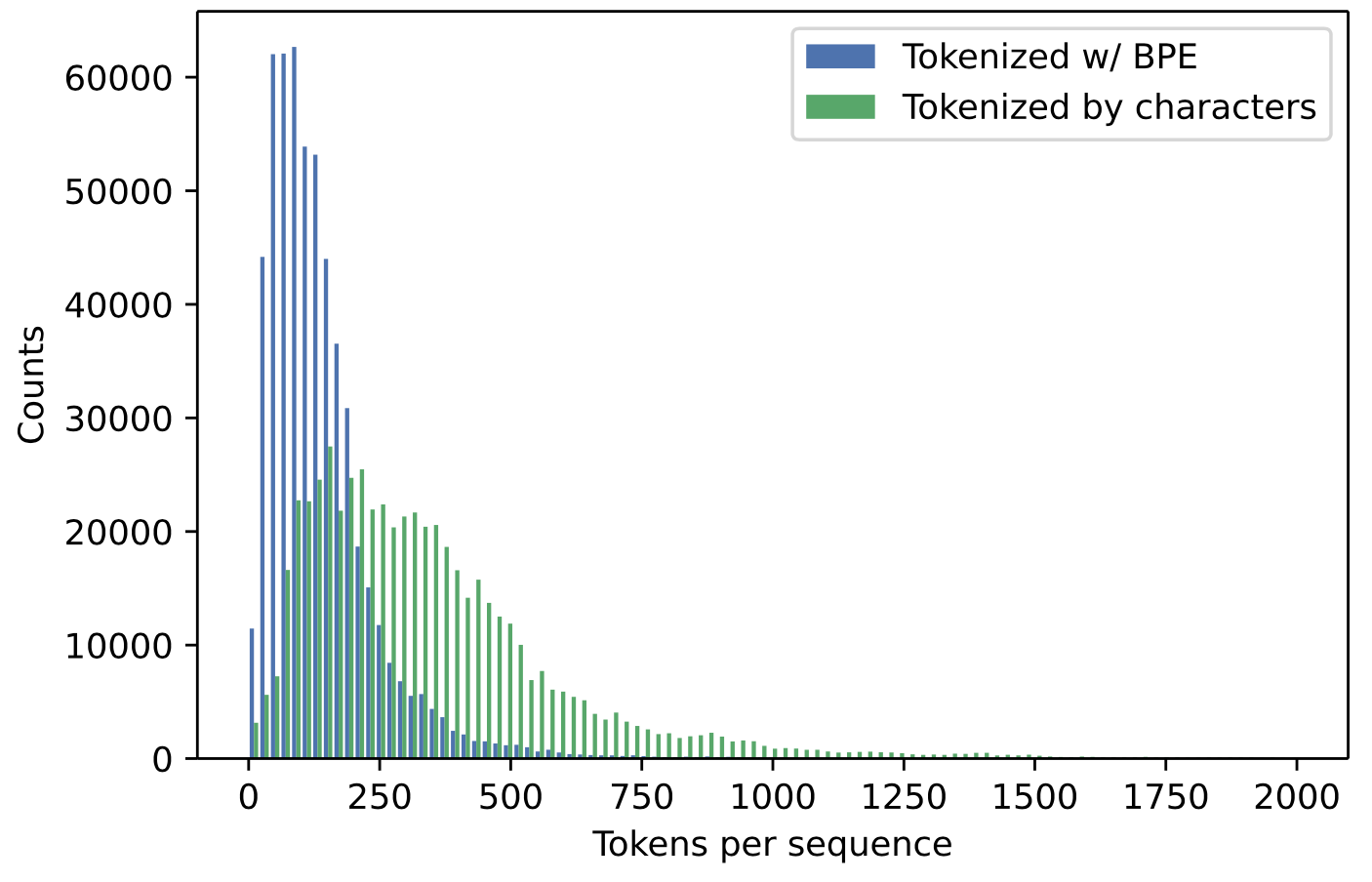}
\caption{BPE vocabulary with 10K tokens captures 98.4\% of SwissProt sequences within the 512 token constraint often used for BERT-like models compared to 82\% if tokenized by single characters.}
\label{fig:bpe_vs_char}
\end{figure}

BPE algorithm creates a predefined size vocabulary from subwords by merging the most frequently occurring subword pairs in a bottom-up fashion, until the vocabulary size is reached \cite{sennrich2015neural}. Amino acid tokens are listed in the order of decreasing value frequency, for an example see \textbf{Appendix Figure \ref{fig:bpe_vocab_examples}}. The most frequent two-letter tokens are "LL", "AA", "AL" and "VL", three letter -- "ALL", "AAL", "LLL". These tokens feature some of the most common amino acids.  Most importantly, as seen in \textbf{Figure \ref{fig:bpe_vs_char}}, BPE more efficiently compresses protein sequences compared to character-level encoding, with an average sequence length down from 360 to 130 tokens -- more than 64\% compression. As a result, only 1.58\% of BPE-tokenized SwissProt sequences exceed the 512 token limit, often seen in BERT-like models, as opposed to a staggering 18.2\% for regular character-level splitting. For a 1,000 token limit, the loss is reduced from 3.3\% to 0.23\%.

Why is the efficient compression so important? First off, most of the studies in the literature discard protein subsequence which exceeds the 500 or 1,0000 amino acid limit, thus losing valuable information. Secondly, for sequence-pair classification tasks, such as PPB prediction, fitting both sequences within the model length constraint is of paramount importance. BPE tokenization for the first time enables us to present two long protein sequences to the Transformer models during the pre-training and fine-tuning stages.

Choosing the vocabulary size parameter $k$ for BPE algorithm was a heuristic process. We have tested $k = {100, 1000, 10000, 30000}$, and observed that at $k=10000$ BPE vocabulary mostly consists of 3 and 4-characters-long tokens, as seen in \textbf{Appendix Figure \ref{fig:bpe_length_distro}}. We postulated that such length distribution should accurately capture the frequent protein subsequence motifs, and be advantageous to the secondary structure prediction problem.

\subsection{Pre-trained embeddings encode amino acid biochemical properties}
We visualized the 768-dimensional embeddings for 21 amino acids by retrieving their corresponding single-letter characters from our 10K subword vocabulary. T-SNE plot in \textbf{Appendix Figure \ref{fig:tsne}} confirms that amino acids cluster according to their charge, hydrophilicity and size, in accordance with \cite{alley2019unified, rao2019evaluating, rives2019biological}. We are still exploring how to visualize the full 10K BPE subword vocabulary.

\subsection{MLM pre-training with protein binding pairs outperforms single sequences}
We present our RoBERTa-like model pre-training results in \textbf{Table \ref{tab:pretraining_results}}. In short, the models are pre-trained on three datasets: SwissProt, Pfam  and String, and two input representations: single sequences (models Pfam, String, SwissProt) or binding sequence-pairs (models String2Seq, StringLF2Seq). Two architectures are pretrained: RoBERTa \cite{liu2019roberta} and Longformer \cite{beltagy2020longformer} which expect up to $T=512$ and $T=2,048$ tokens for model input, respectively.

To test a hypothesis that pre-training on a mixture of binding/non-binding protein sequence-pairs would result in a superior pre-training and downstream fine-tuning task performance (especially for PPB tasks), we have built a novel protein-protein binding dataset from high quality ($>700$) protein binding pairs found in STRING database from 5K organisms and 41M pairs. We selected 2.5M binding and 2.5M non-binding pairs to form our \textit{String2Seq} dataset. The two binders in the sequence pair were separated by \texttt{EOS} and \texttt{SEP} tokens, which in our RoBERTa tokenization scheme share the same token - \texttt{"</s>"}. \textit{String} pretraining dataset consisted of all unique protein sequences found in String2Seq dataset.

As seen in \textbf{Table \ref{tab:pretraining_results}}, String2Seq model had a lower MLM cross-entropy loss than String on Pfam holdout dataset. We hypothesize that the pre-training with a mixture of sequences resulted in learning more accurate secondary and tertiary structure representations, which facilitate protein-protein binding, and by extension, are responsible for the local protein structure. Note, that learning happened exclusively through MLM objective, during AdamW optimization there was no gradient update from sequence-pair relationship information. That is, the model was agnostic whether the two sequences were randomly selected or were strong binders.

\subsection{Sequence-pair pre-trained models outperform single sequence models for fine-tuning tasks}
\input{tables/finetuning_results}

The ultimate test for learning protein language is to evaluate the language model on a downstream task, as \cite{rao2019evaluating} and \cite{strodthoff2019udsmprot} demonstrated that low CE values for MLM task do not always correspond to excellent downstream task performance. We attempted 5 tasks, results for them are summarized in \textbf{Table \ref{tab:finetuning_results}}. Our most shocking finding is the apparent increase in fine-tuning performance after pre-training on 2 sequence pairs only with MLM, without explicit sequence-pair classification objective (like Next Sentence Prediction (NSP) in BERT or Sentence Order Prediction (SOP) in Albert \cite{lan2019albert}).

String2Seq model performed the best for all tasks except for the secondary structure prediction (SSP) and T-cell receptor target prediction (TCR).
For \textbf{localization}, the second best model was String1seq, with accuracy of 0.780. Our model has overtaken the previous result of 0.78 achieved by Bidirectional-LSTM network \cite{almagro2017deeploc}.

String2Seq outscored String1Seq 0.256 to 0.224 (accuracy) on the \textbf{remote homology} task, which maps protein folds to $1,195$ families. This task challenges the model to detect structural similarities among proteins with radically different evolutionary histories. Test sequences are from different protein families than those found in the train set. It is worth noting that the label distribution is extremely unequal, with top 6 most frequent class labels account for 22\% of all data (See \textbf{Appendix Figure \ref{fig:remote_homology_data}}). The previous best result from TAPE study \cite{rao2019evaluating} achieved 0.26 accuracy, with pre-trainend and fine-tuned LSTM, while their BERT implementation was 0.05 lower.

Not surprisingly, String2Seq did extremely well in \textbf{protein-protein binding} (PPB) task, outscoring String1Seq 0.983 to 0.931 (accuracy, 0.445 to 0.251 for F1). This demonstrates that having unlabeled sequence pairs, half of which bind with each other, at training time improves the binding prediction accuracy by 3\%. In part, because it was pre-trained on some of the sequences found in the training set of the task. However, we took care to construct evolutionarily diverse validation and test datasets which do not contain any data used for pre-training the model. 

We are not aware of other methods that propose such pre-training scheme. One study \cite{min2019pre} explored using BERT NSP-like pre-training objective that predicts belonging to the same Pfam family group (3,150 families), but we argue that this task is too easy. Also, we still need to show how the above performance stacks against a model that has both the MLM pretraining objective and the "String Protein-Protein Binding" objective (for further information see Future Directions Appendix \ref{sec:future_nsp}).

Regarding the low performance on \textbf{SSP} task, we were able to test only SwissProt model, that we developed early as one of the baselines. We did not use 10K BPE subword vocabulary, and instead segmented sequences at 20 aa characters. Towards the end of this project we have experienced issues in our sequence annotation testing script. Further work is needed for SSP task (see \ref{seq:future_directions}).

\subsection{Visualizing attention patterns reveals specialized layers in protein LM embedding}
\label{seq:vis_pretrain}

\begin{figure}
\centering
\includegraphics[width=1.0\linewidth]{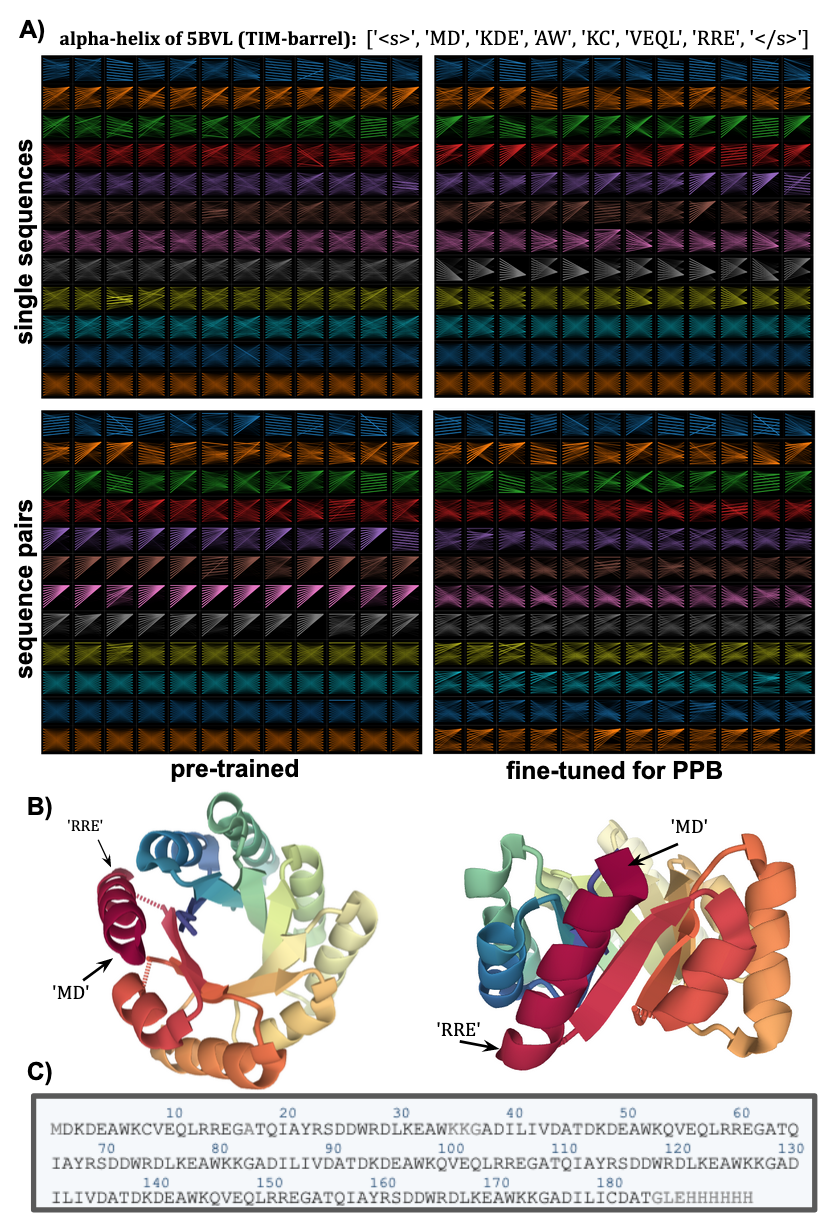}
\caption{Visualizing self-attention between \textit{String} pre-trained and PPB fine-tuned models. \textbf{A)} Unique self-attention patterns visualized with \texttt{BertViz} \cite{vig2019transformervis} for combinations of layers (rows 0-11, top-to-bottom, same color) and attention heads (columns 0-11). We used the 15-aa (7 BPE tokens) alpha-helix of a synthetic protein 5BVL \cite{huang2016novo}). \textbf{B)} 3D views of 5BVL, rendered with \texttt{Mol*} \cite{sehnal2018mol-star}). \textbf{C)} Amino acid sequence for 5BVL, alpha helix of interest corresponds to positions $0-15$.}
\label{fig:alpha_helix}
\end{figure}

Transformers are attention-driven models. We can visualize the multi-head self-attention for the input sequence representation across the layers and heads \cite{clark2019does, rogers2020primer, vig2019transformervis}, which can elucidate if the model learns the syntax and semantics necessary to perform well on prediction tasks.
Initial layers typically encode positional relations, middle -- dependency relations, final - unique and global patterns.

For simplicity purposes, we visualize the layers and attention heads for the first alpha helix of TIM-barrel, a \textit{de novo} designed four-fold symmetry protein \cite{huang2016novo}. As seen in the the two far-left panels on \textbf{Figure \ref{fig:alpha_helix}A}, pre-training on strong protein-protein binders from STRING DB results in drastically different attention patterns. For example, observe the increase in diffuse and unstructured patterns in layers 9, 10 and 11 (teal, blue, orange), or extreme reliance on the (\texttt{BOS}) token, especially in layers 4-7. We speculate that these layers are responsible for secondary structure information.

Most notably, both single and pair-sequence pre-trained models develop attention mechanism that tracks approximately every 4th amino acid, which is responsible for helical secondary structure via hydrogen-bond formation between carboxy and amide groups. In our tokenization scheme this corresponds to attending one or two preceding and proceeding tokens. Attention heads with such behavior were observed in Layers 3-9 in our models. Further attention signatures for other pre-trained models are shown in \textbf{Appendix Figure \ref{fig:pretrained_models}}.

\subsection{Fine-tuning alters attention patterns}
\label{seq:vis_ft}

Fine-tuning has diversified the attention patterns, see \textbf{Figure \ref{fig:alpha_helix}A} right panels. Most notably, sequence-pair pre-training produced embedding layers and heads with extreme focus on the \texttt{BOS} token, as seen in the bottom-left plot layers 4-7. After fine-tuning for protein-protein binding with STRING binders, the attention focused on "VEQL" and "KC" tokens, which have charged Q and K residues, and covalent cystein bond member C. Also, the reliance on BOS and EOS tokens decreases as we fine-tune the model.

Next, we observed that fine-tuning alleviated the non-specific attention as observed in sequence pairs pre-trained model. After fine-tuning, L11 had at least 3 attention heads that were informative and did not focus solely on \texttt{BOS} token.

\section{Conclusions}
In this study we demonstrated that pre-training a RoBERTa language model with MLM objective on a mixture of binding and random protein pairs results in a superior downstream protein classification performance. Next, we compressed the protein sequence space by 64\% by expanding the 20 amino acid character vocabulary to 10K sub-word tokens by BPE algorithm \cite{sennrich2015neural}. Further, we prepared new pre-training and protein-protein binding prediction datasets from STRING database \cite{szklarczyk2019string}. 

Finally to our knowledge, we are the first ones to train a BERT-like model that supports extremely long protein sequences. Although only 3/10,000 proteins are longer than 2,048 tokens (as generated by BPE), those proteins belong to the dark proteome and could benefit from better functional annotation. Also, having a big input representation size is crucial for predicting protein-pair interactions, or even interactions between 3-10 proteins as is common in transcription machinery or when the homo/hetero-protomer complexes assemble.

The language modeling community increasingly advocates for training a large model (in terms of layers and attention heads) on a big dataset for fewer epochs to achieve better downstream results, rather than passing through a small dataset for multiple epochs with a small model \cite{li2020train}. With this dogma in mind, we were somewhat cautious when evaluating the recent protein embeddings \cite{nambiar2020transforming}, that take a similar RoBERTa approach to protein LM, since they trained on a small SwissProt dataset with a shallow and narrow model.


All in all, the study is still largely exploratory and breadth-oriented, and for future studies we aim to focus in depth on tackling the token-level classification tasks such as secondary structure prediction and contact point-prediction which require the model to understand secondary and tertiary structure. We also aim to explore the benefits of adding an explicit protein-protein binding prediction objective. For further future directions see \textbf{Appendix \ref{seq:future_directions}, Future Directions}.

\section*{Broader Impact}
Academic and commercial life scientists continuously characterize protein structures and functionalities. Understanding and then predicting protein properties allows the elucidation of basic biological processes such as gene regulation, and bioengineering of new biologic drugs such as monoclonal antibodies, vaccines and gene editing machinery. Having said that, the commercial businesses, private and government funded research institutes should be continuously scrutinized for potential abuse of proteomics research.
The data used in this study comes from publicly available protein databases that adhere to strict confidentially and anonymization guidelines. All our source code and datasets are publicly available.  We are not aware of any intentions from our research institution to commercialize this research, nor have we applied for a patent.

\bibliographystyle{plain}
\bibliography{neurips_mlsb_2020}

\section*{Appendix}
\subsection*{Fine-tuning dataset preparation}
\label{seq:appendix_finetune_datasets}

\paragraph{Protein-protein binding prediction}
We used the same high confidence binding data from STRING DB. However, this time, the labeled training data was composed of 2M short ($<512$ tokens) and 0.667M long ($>512$ and $\leq2048$ tokens) protein binding pairs, in a 3:1 ratio. Development and test sets had 63.9K and 60.0K sequences, respectively. The dataset was composed of 1:1 binding:non-binding protein pairs \footnote{See \url{scripts.string_prepare_finetuning_dataset}}.

\paragraph{T-cell receptor binding prediction}
The task predicts for a given T-cell Receptor (TCR) the most likely peptide fragments that bind the TCR (peptide fragments from from a larger antigen epitope that's fragmented/digested by the T-cell and subsequently presented on its surface). TCRs here are represented by their very short CDR3 regions which are only 5-20 aa. There are 124,486 TCR and binding epitope pairs, and 13,832 pairs in evaluation dataset.

\paragraph{Remote Homology}
Sequence classification task to assign a protein to one of 1,995 possible folds. Data from \cite{guo2019deep}. Protein fold classes follow Zipf's law, very few folds take up most of 12,311 sequences in training dataset. Top 6 folds take up more than 21 \% of all data.

\paragraph{Solubility}
Binary task to predict if proteins are soluble or not \cite{khurana2018deepsol}. Final training data included 28,972 soluble and 40,448 insoluble proteins, approximately balanced at 7:10 soluble:insoluble ratio.

\input{tables/data_string_finetune.tex}

\input{tables/pretraining_hyperparams}

\begin{figure}
\centering
\includegraphics[width=1.0\linewidth]{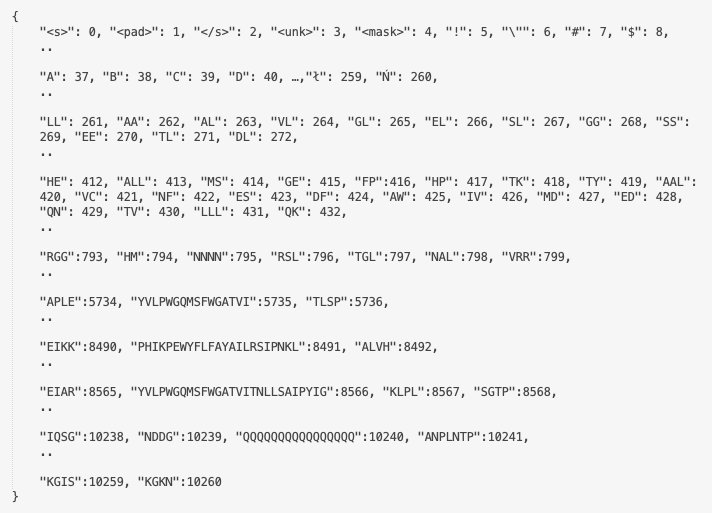}
\caption{BPE vocabulary is a dictionary with token strings arranged in the decreasing frequency serving as keys, and token IDs as the values. Here the most frequent two-letter tokens are "LL", "AA", "AL" and "VL", three letter - "ALL", "AAL", "LLL". Unusually long tokens are observed at 5735, 8491 and 8566 positions}
\label{fig:bpe_vocab_examples}
\end{figure}

\begin{figure}
\centering
\includegraphics[width=1.0\linewidth]{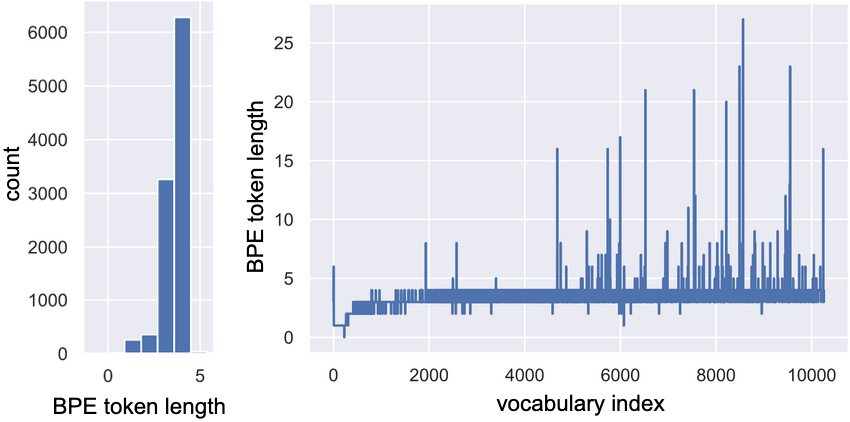}
\caption{Histogram of BPE token lengths. On the left panel we omitted low frequency token lengths (longer than 5). More than 6,000 tokens are 4-characters-long, 3,000 tokens consist of 3 characters, and only less than 200 tokens are 1 or 2 characters-long. Some rare unusually long tokens appear on the right figure, such as 23 character \texttt{PHIKPEWYFLFAYAILRSIPNKL} and 27 character \texttt{YVLPWGQMSFWGATVITNLLSAIPYIG} at indices 8491 and 8566 respectively.}
\label{fig:bpe_length_distro}
\end{figure}

\begin{figure}
\centering
\includegraphics[width=1.0\linewidth]{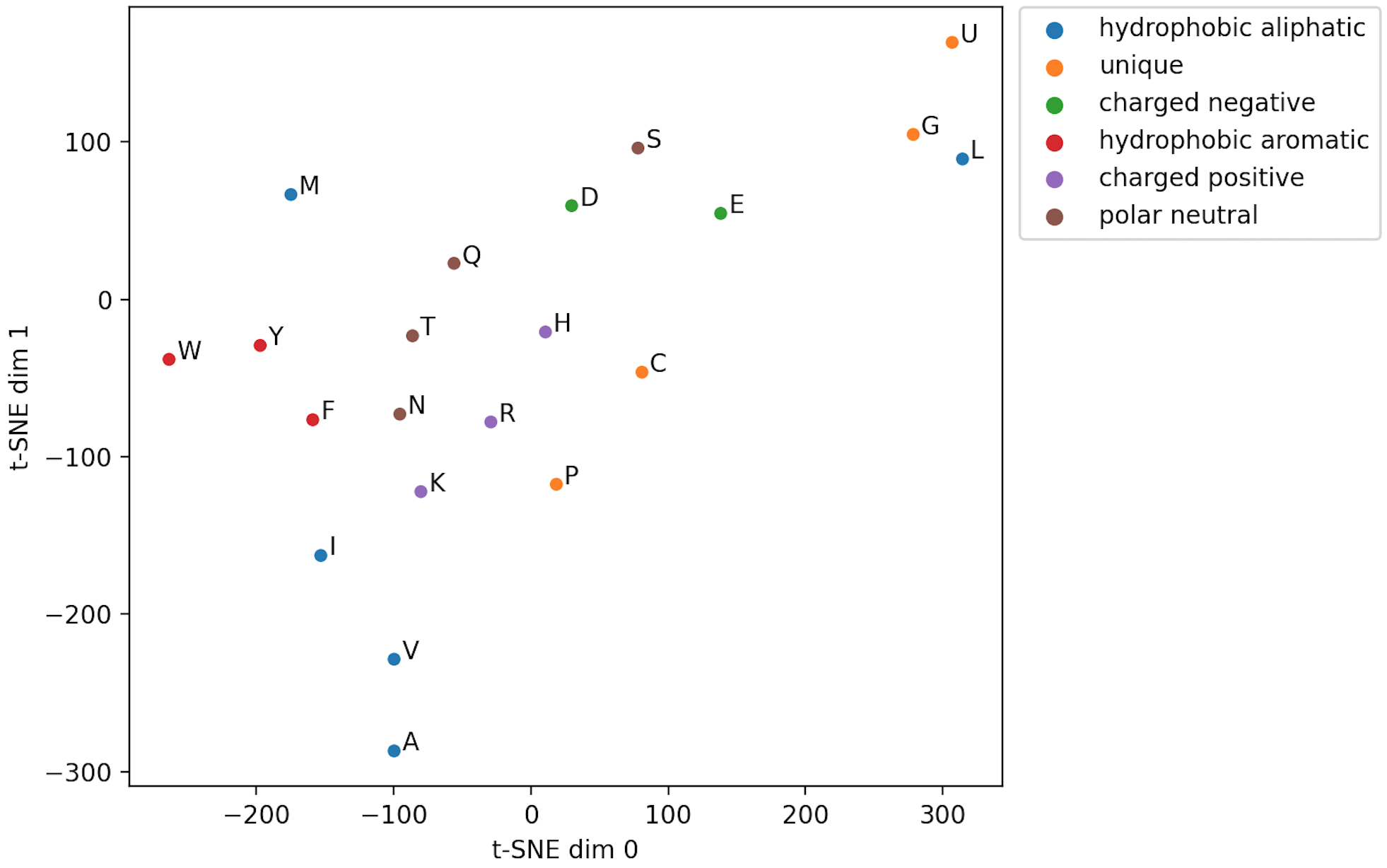}
\caption{t-SNE visualization of 768-dimensional embedding space for 21 single character BPE tokens that encode amino acids. Parameters for t-SNE compression: learning rate: 40, perplexity: 5}.
\label{fig:tsne}
\end{figure}

\begin{figure}
\centering
\includegraphics[width=0.9\linewidth]{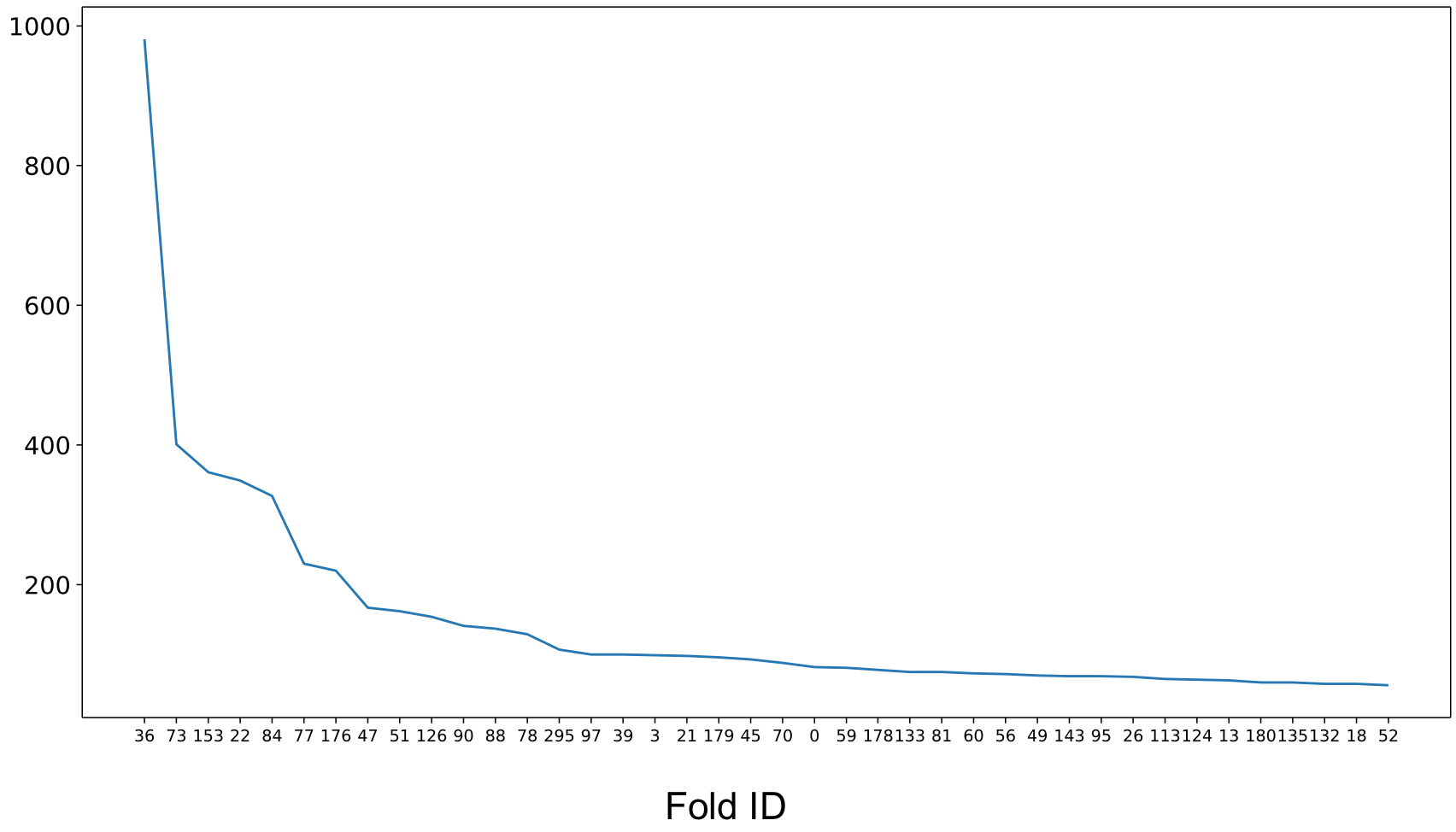}
\caption{Top 40 most common protein fold classes among 12,311 proteins in the remote homology dataset.}
\label{fig:remote_homology_data}
\end{figure}

\subsection*{Additional attention mechanism analysis}

\begin{figure}
\centering
\includegraphics[width=1.05\linewidth]{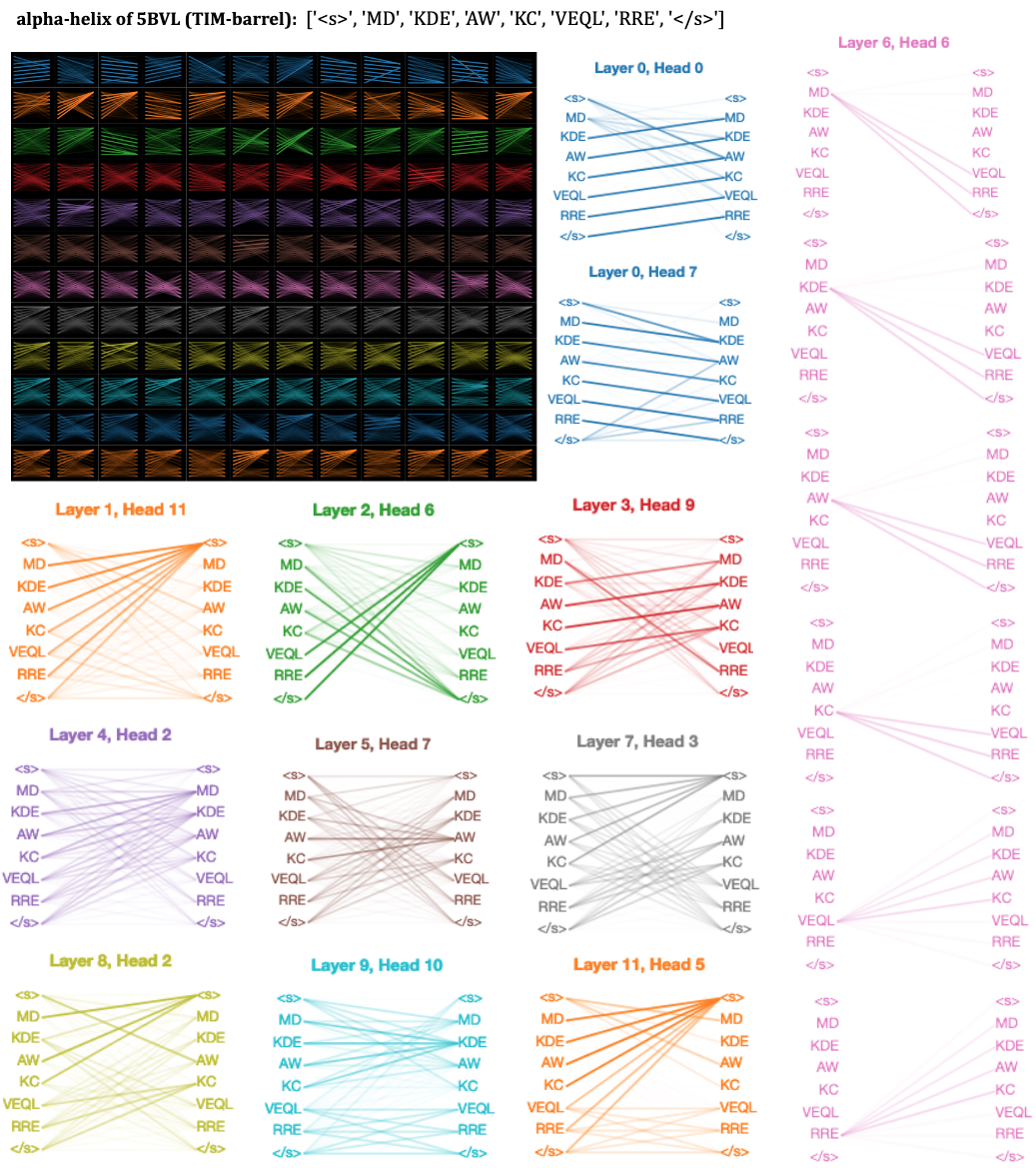}
\caption{In-depth look at self-attention patterns of \textit{String1Seq} model after fine-tuning for protein binding prediction. Showing alpha-helix of a synthetic protein 5BVL \cite{huang2016novo}.
Self-attention for 8 tokens shown for all combinations of model layers (rows 0-11, top-to-bottom, same color) and attention heads (columns 0-11). Visualized with \texttt{BertViz} \cite{vig2019transformervis}.}
\label{fig:alpha_helix_zoom}
\end{figure}

Here we give a glimpse into the attention mechanisms of our single-sequence String DB pretrained model, String1Seq, \ref{fig:alpha_helix_zoom}. Initial Layer 0 (L0) exhibits attention to previous and next tokens. Layer L1 mostly, focuses on BOS token, while L2 on a mixture of BOS and EOS tokens. One general pattern we observed is that the reliance on BOS and EOS tokens decreases as we fine-tune the model and as we progress upwards through its layers. For example, attention on previous tokens is strong in L3H9, which is probably instrumental for the model's secondary structure understanding, but unlike in L1-2, BOS and EOS tokens are largely ignored.

Starting with L3-4 we observe several attention modes per head, while in the previous layers one mode was always dominating. For instance, in L4H2 there is a strong self attention towards the first half of sequence, \texttt{['MD', 'KDE', 'AW', 'KC']}, while in L4H11, first half of sequence focuses on the next token and second half of the sequence on the first half. Similarly L5 has a mixture of heads that look onto the last tokens, or focus on \texttt{['AW', 'KC']}. We show all token-token attention patterns for L6H6 in the right-most panel, interestingly, the first 4 protein tokens exclusively focus on the last two tokens \texttt{['VEQL', 'RRE']}, while the last two tokens focus only on \texttt{['MD', 'KDE', 'AW', 'KC']}. This uncovers another pattern, of learning to pay attention separately to the two halves of the sequence. L7-9 demonstrate multiple modes in most of their attention heads. Interestingly, in L9H10, we see the opposite of L6H6 - the set of the first four tokens and the final two tokens pay attention only to their respective subsets. Layer 11 is dominated by this pattern, except that the first half pays attention to BOS token. 

\begin{figure}
\centering
\includegraphics[width=1.0\linewidth]{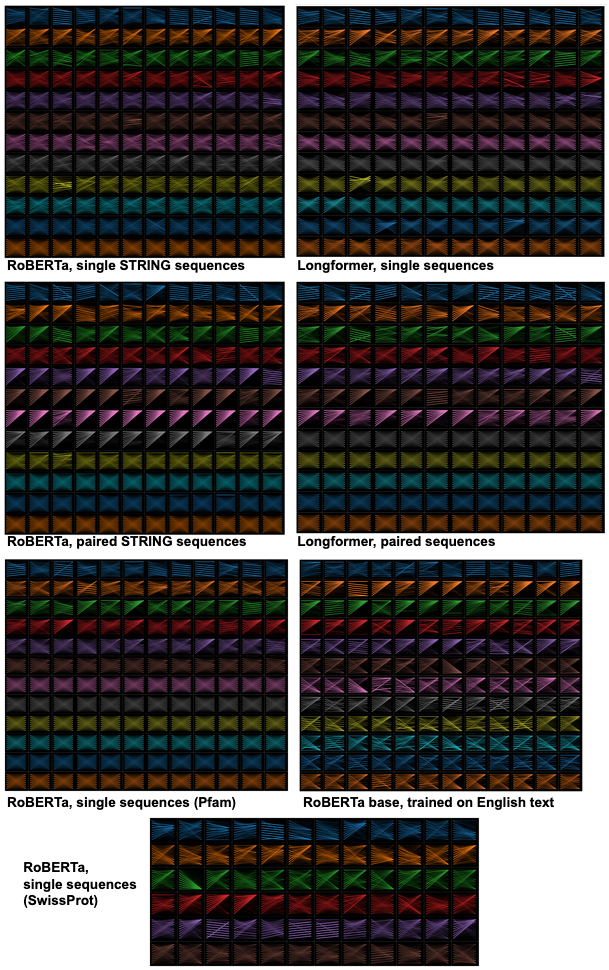}
\caption{Attention mechanisms across layers and self-attention heads for other pre-trained models. Visualized with \texttt{BertViz} \cite{vig2019transformervis}.
}
\label{fig:pretrained_models}
\end{figure}

\section*{Future directions}
\label{seq:future_directions}

As is common in science, by the end of this study there are more unanswered questions than at the beginning. Here we present a few directions where the project could be extended.

\subsection*{Secondary structure prediction on Byte-Pair encoded sequences}
\label{seq:sec_struct_future}

\begin{figure*}[h]
\centering
\includegraphics[width=0.9\linewidth]{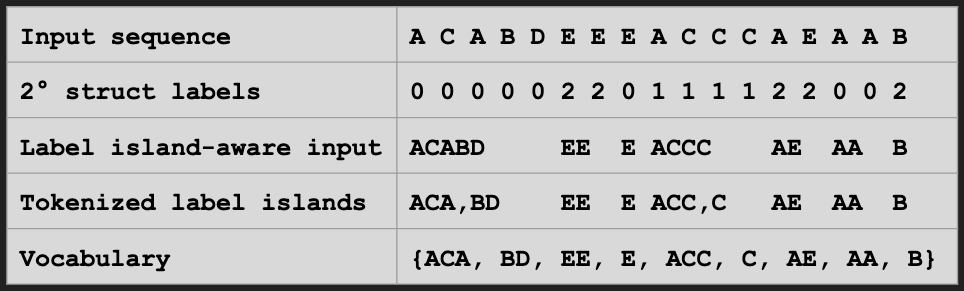}
\caption{Label island continuity problem. }
\label{fig:token_islands}
\end{figure*}

\begin{figure*}[ht]
\centering
\includegraphics[width=0.8\linewidth]{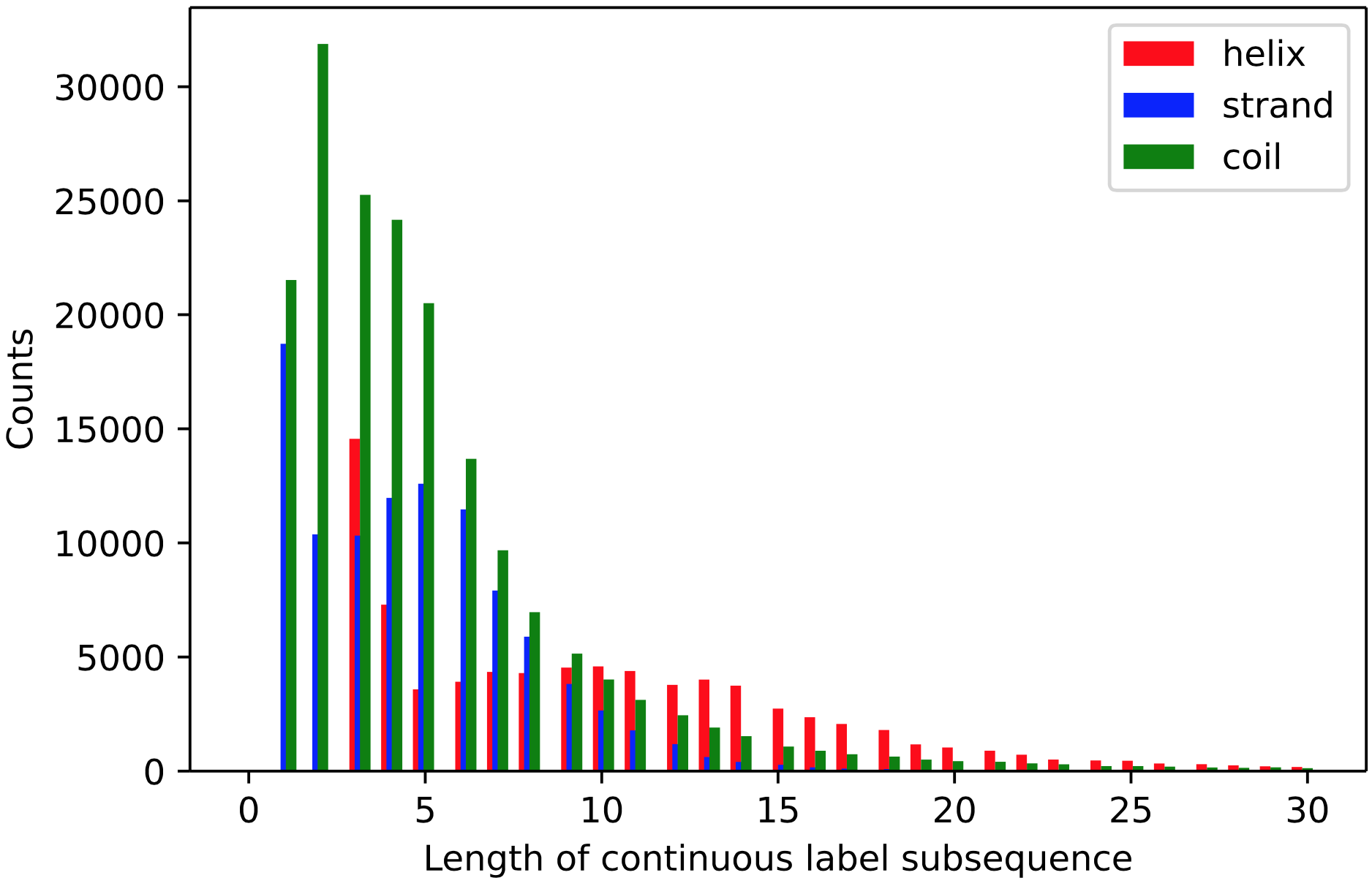}
\caption{Secondary structure labels often form 2-7 amino-acid-long sub-sequences, which we call \textit{label islands}.}
\label{fig:second_struct_island_lengths}
\end{figure*}

So far we have used a 20 amino acid vocabulary when doing a 3-class secondary structure predictions. Training a protein model with 20-25 amino acid vocabulary is straightforward, we treat each character as a token which has a single corresponding label, and predict new labels in the test set. But what if we have 10k tokens from BPE, and assume that a set of longer, multi-character BPE tokens will carry more structural information to predict secondary structure. To utilize the full 10k vocabulary we have to overcome a problem: input sequence tokens (after BPE encoding) will break the continuity of labels. See this problem illustrated in \textbf{Figure \ref{fig:token_islands}}.
Secondary structure labels often form 2-7 amino-acid-long sequences, which we call label islands, see \textbf{Figure \ref{fig:second_struct_island_lengths}}. We are proposing to explore pre-tokenizing all the training data with BPE tokenizer and creating a new label-island aware vocabulary, and prune the BPE vocabulary by removing the tokens that do not appear in the new vocabulary. Then during evaluation, we would hope that the test sequences are correctly parsed, and label continuity is preserved after tokenizing with the new pruned vocabulary. Alternatively, we could split the sequence into tri-grams.


\subsection*{Searching for a suitable Next Sentence Prediction objective in the protein domain}
\label{sec:future_nsp}
We propose further exploration if and what supplementary intrinsic pretraining objective (besides the masked language modeling, or MLM) could be beneficial for the secondary structure and PPI prediction fine-tuning tasks.
For example, Min et al (2020) \cite{min2019pre} proposed the Same Familiy Prediction objective. However as we saw in \cite{rao2019evaluating, rives2019biological}, clustering proteins into Pfam families is a relatively easy task with $>90\%$ accuracies. Therefore, would propose using Protein-Protein Binding (PPB) objective, because the model would have to explicitly learn the feature space responsible for detecting binding epitopes. The model implementation is quite simple, we could simply apply a binary mask on our input, that signifies which tokens belong to which sequence, as is done in BERT \cite{devlin2018bert} model.


\subsection*{Investigate new uses for extremely large model input space (2,048 tokens) made available by Longformer pre-training}
We estimate that a model with $T=2048$ can accomodate more than 99.99\% protein sequences. In theory, if we assume that self-attention can fully understand protein secondary and tertiary structure, with such a large model we could predict the binding between two (1,022 each), four (511) or eight (255) proteins. But first, we still have to validate the Longformer on all downstream tasks.

\subsection*{Formulate TCR binding prediction as a Question Answering task}
T-cell receptor binding prediction is extremely hard to model. We propose to introduce a new protein sequence task akin to question answering in NLP. We could train on full TCR sequences, not just CDR3, with an additional objective which determines the exact span where the short CDR3 lies in the long TCR sequence.

\subsection*{Visualization and quantification of binding-site predictions}
We assume that by training on protein sequence pairs, the self-attention heads learn across the layers which subsequence patterns lead to the two-sequence binding and recognize what patterns constitute the binding sites. By visualising combinations of $L$ layers and $A$ self-attention heads, we try to retrieve the correct binding sites. Instead of the simple binding pair visualization by BertViz, we are curious to utilize RXNMapper \cite{Schwaller2020Unsupervised} to discover the binding sites, and give probabilities for binding site prediction.

\subsection*{Novel subword encoding strategies}
Once the BPE subword vocabulary, or the merges table, is established, BPE segments any given word in a deterministic fashion. BPE could be further improved by randomly removing subwords from the vocabulary before each training epoch \cite{provilkov2019bpe}. As the BPE merges table is being assembled, we can "regularize" the table by randomly dropping out subword merge pairs with probability $p=0.1$.

According to \cite{bostrom2020byte}, Unigram language modeling \cite{kudo2018sentencepiece}, albeit more computationally expensive, is more advantageous than BPE when pre-training language models. Whereas BPE creates new subword tokens while $|V| < k$, ULM algorithm starts with a superset of the final vocabulary, and while $|V| > k$ ($V$ is a set of all substrings occuring more than once in the set of strings $D$) proceeds to prune out tokens, by inferring the language model parameters $\theta$. Using Viterbi algorithm, the optimal segmentation is determined (this segmentation has maximal likelihood with inferred $\theta$).

\end{document}

%% file: tables/pretraining_results.tex
\begin{table*}
\begin{center}
\begin{tabular}{lcccccc}
\toprule
\textbf{Datasets}  & Pfam & String & String2Seq & StringLF & StringLF2Seq & SwissProt \\
\midrule 
CE train & 5.01 & 4.70 & 4.59 & 4.91 & 5.16 & 4.58 \\
CE val & 5.81 & 5.83 & 5.22 & 6.52 & 6.30 & 4.49 \\
CE Pfam holdout & 7.40 & 7.72 & \textbf{6.41} & 8.03 & 6.75 & 6.56 \\
Train data size & 31M & 10M & 5M & 9.53M & 4.76M & 504K \\
Eval data size & 100K & 150K & 150K & 150K & 150K & 56K \\
Batch size & 128 & 128 & 128 & 16 & 16 & 128 \\
Train Epochs & 2 & 3.4 & 6.5 & 0.030 & 0.070 & 45.3 \\
Final LR & 1.5e-5 & 3.8e-7 & 7.4e-8 & 6.6e-5 & 6.2e-5 & 5.5e-5\\
Training Steps & 561K & 300K & 321K & 66K & 63K & 178K \\
Round 1 steps & 389.6K & 144K & 158K &  &  & \\
Round 2 steps & 171.5K & 155K & 163K &  &  & \\

\bottomrule
\end{tabular}
\end{center}
\caption{Hyperparameters and language model pre-training results for different pre-training datasets. Cross Entropy (CE) loss is a natural log of perplexity. \textbf{Disclosure}: Round 2 for String1Seq and String2Seq conducted because the LR was not decreasing fast enough, LR was 9.84e-5 and 9.64e-5 respectively. In round 2 the learning rate was linearly decreased to 0. Pfam for round 1 was trained with a linearly decreasing schedule, but accidentally set for expected 100 epochs instead of 5, only 1.6 of which were completed. Round 2 continued from round 1 checkpoint with 2 expected epochs, 0.4 of which were completed, with the final LR 1.48e-5. Pfam holdout set consists of 44,311 sequences.}
\label{tab:pretraining_results}
\end{table*}

%% file: tables/data_all_finetune_tasks.tex
\begin{table*}
\begin{center}
\begin{tabular}{lcccccc}
\toprule
\bf Tasks  & \bf Localization & \bf Solubility & \bf Homology & \bf PPB & \bf TCR & \bf SSP \\
\midrule 
Reference & \cite{almagro2017deeploc} & \cite{khurana2018deepsol} & \cite{chandonia2019scope} & \textit{ours and} \cite{szklarczyk2019string} & \textit{Weber2020\textbf{*}} & \cite{klausen2019netsurfp} \\
Train & 9,977 & 62,478 & 12,311 & 2,662,711 & 124,486 & 9,279\\
Dev & 1,107 & 6,942 & 734 & 63,589 & 13,832 & 2316\\
Test & 1,000 & 2,001 & 2,000 & 150,000 & 10K & 581\\ 
\bottomrule
\end{tabular}
\end{center}
\caption{
Fine-tuning datasets. PPB is Protein-Protein Binding, TCR is T-cell receptor and its epitope binding, SSP is secondary structure prediction. \textbf{*}located at: \url{https://ibm.ent.box.com/v/paccmann-proteomics-data/folder/126008768303}.
}
\label{tab:data_all_finetune_tasks}
\end{table*}

%% file: tables/finetuning_results.tex
\begin{table*}[t]
\begin{center}
\begin{tabular}{lcccccc}
\toprule
\bf Tasks  & \bf Localization & \bf Solubility & \bf Homology & \bf PPB & \bf TCR & \bf SSP \\
\midrule 
Result (accu) & \textbf{0.793} & 0.583 & 0.256 & \textbf{0.983} & \textbf{0.729}* & 0.669* \\
SOTA result & 0.78 & 0.77 & 0.26 & NA & NA & 0.791**\\
Classes & 10 & 2 & 1,195 & 2 & 2 & 3 \\
LR & 5e-5 & 5e-5 & 5e-5 & 3e-5 & 5e-5 & 1e-5\\
Batch Size & 16 & 8 & 16 & 32 & 32 & 8\\
Epochs & 15 & 15 & 15 & 2 & 5 & 5\\
Optim steps & 9,360 & 58,575 & 11,550 & 166K & 19,455 & 5,800\\
\bottomrule
\end{tabular}
\end{center}
\caption{
Fine-tuning results, reported in accuracy. All of the best results achieved with String2Seq model, unless specified by "*". *SwissProt; **3-class secondary structure prediction, using SwissProt pretaining, SoTA result from 12-layer BERT from \cite{rives2019biological}, their best 34 layer model had accuracy of 0.89. We ran our fine-tuned models against the test sets only once. Development runs with a unique set of parameters were not replicated. Hyperparameters for best performing models are shown. 
}
\label{tab:finetuning_results}
\end{table*}

%% file: tables/data_string_finetune.tex
\begin{table*}
\begin{center}
\begin{tabular}{lccc}
\toprule
\bf STRING split & \bf Train & \bf Dev & \bf Test\\
\midrule
loose total avail. & 54.4M (0.89) & - & 22.1M \\
loose used & 4.00M (0.75) & 0.200M & 0.600M \\

strict total avail. & 54.4M (0.89) & - & 248K (0.37) \\
strict used &  2.67M (0.75) & 63.9K & 60.0K \\

\bottomrule
\end{tabular}
\end{center}
\caption{Preparing STRING dataset for model fine-tuning. In parentheses we show what fraction of sequence pairs after tokenization are less than 512 tokens-long (sequences $A$ and $B$, plus special tokens), the remainder is between 512 and 20148 tokens.}
\label{tab:data_string_finetune}
\end{table*}

%% file: tables/pretraining_hyperparams.tex
\begin{table*}[t]
\begin{center}
\begin{tabular}{lcccc}
\toprule
\textbf{Hyperparameter}  & Pfam & String1(2)Seq & StringLongformer1(2)Seq & SwissProt \\
\midrule 
Number of Tokens & 512 & 512 & 2048 & 512 \\
Number of Layers & 12 & 12 & 12 & \textbf{6}\\
Hidden size & 768 & & &\\
FFN inner hidden size & 3072 & & &\\
Attention heads & 12 & & &\\
Attention head size & 64 & & &\\
Dropout & 0.1 & & &\\
Attention Dropout & 0.1 & & &\\
Warmup Steps & 4K & 20K & 512 & 5K \\
Peak Learning Rate & 1e-4 & 1e-4 & 7e-5 & 1e-4\\
Weight Decay & 0.01 & & &\\
Total Batch Size & 128 & 128 & 16 & 128\\
Per GPU Batch Size & 16 & 16 & 1 & 16\\
Number GPUs & 4 & & & \\
GA Steps & 2 & 2 & 4 & 2 \\
Tokens per Batch & 65,536 & & & \\
FP16 & yes  & & & \\
AdamW $\epsilon$ & 1e-6  & & & \\
AdamW $\beta_1$ & 0.9  & & & \\
AdamW $\beta_2$ & 0.98  & & & \\

\bottomrule
\end{tabular}
\end{center}
\caption{
Hyperparameters for pretraining Protein RoBERTa. If a hyperparameter showed only once, assume it is applied to all models.}
\label{tab:pretraining_hyperparams}
\end{table*}